\documentclass[12pt]{article}
\def\be{\begin{equation}}
\def\ee{\end{equation}}
\def\bea{\begin{eqnarray}}
\def\eea{\end{eqnarray}}
\def\s{\sigma}

\def\om{\omega}
\def\pr{\prime}
\begin{document}

\title{Strings in meson and baryon physics}
\author{G. S. Sharov \\
\small\it
Tver State University,
Sadovyj per., 35, 170002, Tver, Russia}

\maketitle

\begin{abstract}

The relativistic string with massive ends (the meson model) and four various
string baryon models: $q$-$qq$,  $q$-$q$-$q$, Y and $\Delta$
are considered. In particular, the rotational motions for all these systems
are applied to describing the leading Regge trajectories,
and also the classical dynamics beyond the usual rectilinear string rotations
is studied. For the string meson model the two types of quasirotational motions
 (disturbances of the planar uniform rotations)
are obtained. They are oscillatory motions in the form of stationary
waves in the rotational plane and in the orthogonal direction.
The analysis of the stability problem for the rotational motions
for all mentioned string configurations shows that these motions
for the Y and $q$-$q$-$q$ models are
unstable on the classical level.

\end{abstract}


\bigskip

In various string models of hadrons the relativistic string simulates
the strong interaction between quarks at large distances and the QCD
confinement mechanism  \cite{BN}. One of the starting points for strings
in particle physics was the natural explanation of linearly growing
Regge trajectories for orbitally excited hadron states, where the
Regge slope $\alpha'=1/(2\pi\gamma)$ is simply connected with the
string tension $\gamma$.

The string model of the meson is geometrically obvious,
on the classic level it is the
relativistic string with massive ends shown in Fig.~1 (a) \cite{BN}.
But for the baryon
we have four string models with different topology \cite{AY}:
(b) the meson-like quark-diquark model $q$-$qq$ \cite{Ko},
(c) the linear configuration $q$-$q$-$q$ \cite{lin},
(d) the ``three-string" model or Y configuration \cite{CollinsPY},
and (e) the ``triangle" model or $\Delta$ configuration \cite{Tr}.

\begin{figure}[bh]
\unitlength=0.8mm
\begin{center}
\begin{picture}(154,40)
\thicklines
\put(50,35){\line(1,0){48}}
\put(50,35){\circle*{2}}\put(98,35){\circle*{2}}
\put(46,30){$q$}\put(100,30){$\overline q$}\put(66,38){(a)}
\put(7,15){\line(1,0){25}}
\put(7,15){\circle*{2}}\put(32,16){\circle*{2}}\put(32,14){\circle*{2}}
\put(6,10){$q$}\put(29,9){$qq$}\put(11,20){(b)}
\put(47,15){\line(1,0){28}}
\put(47,15){\circle*{2}}\put(61,15){\circle*{2}}\put(75,15){\circle*{2}}
\put(46,10){$q$}\put(60,10){$q$}\put(74,10){$q$}\put(52,20){(c)}
\put(97,15){\line(0,-1){12}}\put(97,15){\line(-2,1){10}}
\put(97,15){\line(2,1){10}}
\put(97,3){\circle*{2}}\put(87,20){\circle*{2}}\put(107,20){\circle*{2}}
\put(93,3){$q$}\put(83,17){$q$}\put(108,17){$q$}\put(93,22){(d)}
\put(115,5){\line(1,0){24}}\put(115,5){\line(2,3){12}}
\put(127,23){\line(2,-3){12}}
\put(115,5){\circle*{2}}\put(139,5){\circle*{2}}\put(127,23){\circle*{2}}
\put(111,4){$q$}\put(141,4){$q$}\put(123,24){$q$}\put(136,20){(e)}
\end{picture}\end{center}
\caption{String models of the meson (a) and the baryon (b)\,--\,(e).}
\end{figure}

The quark-diquark configuration was studied better (in comparison with other
string baryon models) because of its simplicity and similarity with the meson
model. But the linear $q$-$q$-$q$ configuration was less popular because it
was supposed to be unstable with respect to transformation into the
quark-diquark one \cite{Ko}. However this problem was studied quantitatively
only recently in Ref.~\cite{lin}, where we showed that the
classic rotational motions of this model are unstable indeed, but the system
doesn't transform into the $q$-$qq$ one. The stability problem of the classic
rotations for all other baryon model was studied in Ref.~\cite{stabPRD}.
This was surprising, but the Y configuration appeared to be unstable
too.

For the three-string model or Y configuration the massless variant was better
investigated \cite{AY,CollinsPY} in comparison with $q$-$q$-$q$ or $\Delta$
ones (for other models the massless case is trivial non-baryon open or closed
string), but even for the massless three-string
the dynamics remains too complicated.
The delay with study of the last baryon model ``triangle" was
also connected with some difficulties in its dynamics.

The problem of choosing the most adequate string baryon model among
the four mentioned ones has not been solved yet, because
All these models can work in the particle physics under certain
assumptions \cite{4B}. All models have a some degree of the QCD
motivation but without explicit preferences. In particular,
different authors in the frameworks the QCD-based baryon Wilson loop operator
approach give some arguments in favour of the Y or
the $\Delta$ configuration  \cite{KalashCorn}.

For all mentioned string hadron models the dynamics may be described by
the action \cite{BN,4B}
\be
S=-\gamma\int\limits_\Omega\sqrt{-g}\,d\tau\,d\s-
\sum_{i=1}^N m_i\int\sqrt{\dot x_i^2(\tau)}\,d\tau,
\label{S}\ee
where $\gamma$ is the string tension,
$m_i$ are masses of $N$ material points (modelling quarks, antiquarks
or diquarks), $N=2$ for the meson-like models (a), (b)
in Fig.~1 and $N=3$ for others,
$X^\mu(\tau,\s)$ are coordinates of a string point in Minkowski
space with signature $+,-,-,\dots$,
$g=\dot X^2X'{}^2-(\dot X,X')^2$,
$\dot X^\mu=\partial_\tau X^\mu$, $X^{\pr\mu}=\partial_\s X^\mu$,
the domain $\Omega$ mapping into the world surface is bounded by the
inner lines $\s=\s_i(\tau)$ of the material points, $\Omega$ is different for
different configurations \cite{stabPRD,4B},
$\dot x_i^\mu=\frac d{d\tau} X^\mu(\tau,\s_i(\tau))$, $c=1$.

For the triangle configuration the string is closed (but it is not smooth)
so we use the closure condition \cite {Tr}
\be
X^\mu(\tau,\s_0(\tau))=X^\mu(\tau^*,\s_3(\tau^*)).\label{cl}\ee
Here $\s=\s_0(\tau)$ and $\s=\s_3(\tau)$
describe the trajectory of the same quark.

For the three-string baryon model in three parametrizations
$X_i^\mu(\tau_i,\s)$ of the three world sheets the different ``time-like"
parameters $\tau_i$ \cite{stabPRD} are connected at the junction world line
$\tau_2=\tau_2(\tau)$, $\tau_3=\tau_3(\tau)$, $\tau_1\equiv\tau$. So
the general form of the junction condition is
\be
X_1^\mu\big(\tau,0\big)=X_2^\mu\big(\tau_2(\tau),0\big)=
X_3^\mu\big(\tau_3(\tau),0\big).
\label{junc}\ee

The equations of motion and the boundary conditions at the
trajectories of massive points for all the models
result from action (\ref{S}).
If we choose coordinates $\tau$, $\s$ (it is possible for all the models
\cite{BN,Tr,stabPRD}) satisfying the orthonormality conditions or
conditions of conformally flat induced metric
\be
\dot X^2+X'{}^2=0,\qquad(\dot X,X')=0,\label{ort}\ee
the equations of motion become linear
\be
\ddot X^\mu-X''{}^\mu=0,\label{eq}\ee
but the boundary conditions for the massive point at an end
\be
m_i\frac d{d\tau}U^\mu_i(\tau)\pm\gamma
\big[X'{}^\mu+\s_i'(\tau)\,\dot X^\mu\big]
\Big|_{\s=\s_i}=0,\quad\; U^\mu_i(\tau)=
\frac{\dot X^\mu+\s_i'X^{\pr\mu}}{|\dot X+\s_i'X'|}\bigg|_{\s=\s_i}\!\!
\label{qq}\ee
or in the middle point (for the models $q$-$q$-$q$ or $\Delta$)
\be
m_i\frac d{d\tau}U_i^\mu(\tau)
-\gamma\big[X'{}^\mu+\s_i'(\tau)\,\dot X^\mu\big]
\Big|_{\s=\s_i+0}
+\gamma\big[X'{}^\mu+\s_i'(\tau)\,\dot X^\mu\big]
\Big|_{\s=\s_i-0}\!\!=0,
\label{qqq}\ee
remain essentially nonlinear.
These massive points make the models much more realistic
but they bring additional nonlinearity and (hence) a lot of problems
with quantization of these models.

The well known exact solution of Eq.~(\ref{eq}) satisfying orthonormality
(\ref{ort}) and all the boundary conditions
describes the rotational motions of the meson-like, $q$-$q$-$q$ and Y string
configurations (flat uniform rotations of the rectilinear string segments)
and may be represented in the form \cite{Ko,stabPRD,4B}:
\be
X^0\equiv t=\tau,\qquad
X^1+iX^2=\om^{-1}\sin\om\s\cdot e^{i\om\tau}.
\label{sol}\ee
Here  $\om$ is the angular velocity, $\s_i={}$const.

For the linear $q$-$q$-$q$ configuration the middle quark is at rest at
$\s=\s_2=0$ and for the three-string model 3 rotating rectilinear string
segments joined in the rotational plane at the angles 120${}^\circ$
\cite{AY,CollinsPY}.

Rotational motions for the baryon model ``triangle" have
the form \cite{Tr,4B}
\be\textstyle
X^0=\tau-\frac TD\s,\qquad
X^1+iX^2=u(\s)\cdot e^{i\om\tau}.
\label{soltr}\ee
This exact solution of Eq.~(\ref{eq}) satisfies the
orthonormality (\ref{ort}), closure (\ref{cl}), boundary (\ref{qqq})
conditions and describes an uniformly rotating closed string (curvilinear
triangle) composed of three segments of a hypocycloid.
In Eq.~(\ref{soltr}) $u(\s)=A_i\cos\om\s+B_i\sin\om\s$,
$\s\in[\s_i,\s_{i+1}]$, the six complex constants $A_i,B_i$
and the real constants $\s_i$, $D=\s_3-\s_0$, $V_i^2$, $\tau^*-\tau=T$
are connected by the set of relations \cite{Tr,4B,InSh}.

The energy $E$ and angular momentum $J$ of the states (\ref{sol}),
(\ref{soltr}) are \cite{Tr,4B}
\be
E=E_{st}+\sum_{i=1}^N\frac{m_i}{\sqrt{1-v_i^2}}+
\Delta E,\quad
J=\frac1{2\om}\bigg[E_{st}
+\sum_{i=1}^N\frac{m_iv_i^2}{\sqrt{1-v_i^2}}\bigg]+S,
\label{EJ}\ee
where $v_i$ are the quark velocities,
$E_{st}=\gamma\om^{-1}\arcsin v_i$ for the motions (\ref{sol}) and
$E_{st}=\gamma D(1-T^2/D^2)$ for the triangle states (\ref{soltr}).
The quark spins with projections $s_i$ (S=$\sum_{i=1}^Ns_i$)
are taken into account, in particular, as the spin-orbit correction
$\Delta E=\Delta E_{SL}=\sum\limits_i\beta(v_i)(\vec\om\vec s_i)$
to the energy of the classic motion. Here we use
$\beta(v_i)= 1-(1-v_i^2)^{1/2}$ for this correction \cite{4B,InSh}.

The expression (\ref{EJ}) for all string hadron models in Fig.~1
describes quasilinear Regge trajectories with the similar ultrarelativistic
behavior ($v_i\to1$):\newline
$
J\simeq\alpha'E^2-\nu E^{1/2}\sum_{i=1}^Nm_i^{3/2}
+\sum_{i=1}^Ns_i\big[1-\beta(v_i)\big].
$
Here the slopes are different: $\alpha'=(2\pi\gamma)^{-1}$ for the meson-like
models, $\alpha'=\frac23(2\pi\gamma)^{-1}$ for the Y and
$\alpha'=n(n^2-k^2) ^{-1}(2\pi\gamma)^{-1}$ for the ``triangle" \cite{4B,InSh}.

The parent Regge trajectories for the $N$, $\Delta$ and strange
baryons may be described with using all string baryon models under following
assumptions: $\gamma=\gamma_{q-qq}=0.175$ GeV${}^2$ the effective
tension for the Y and ``triangle"  is to be different
$\gamma_Y=\frac23\gamma$, $\gamma_\Delta=\frac38\gamma$.
Under these assumptions and the effective quark masses $m_u=m_d=130$ MeV,
$m_s=270$ MeV the mentioned baryonic trajectories and also the
Regge trajectories for the light and strange mesons are well described
\cite{4B,InSh}.

The model ``triangle" has a set of topologically different
solutions (\ref{soltr}) \cite{Tr} numerated by
the integer numbers
$n=\lim\limits_{m_i\to0}D/(\s_1-\s_0)$, $k=n\lim\limits_{m_i\to0}T/D$.
The states with $n=3$, $k=1$ (``simple states" \cite{Tr})
are stable \cite{stabPRD} and they are used here for describing the Regge
trajectories.
Other values $n$, $k$ correspond to the exotic states with some string
points moving at the speed of light \cite{Tr}. These states are unlikely
be physical (except for possible describing glueballs and
hybrids) but they may act as physical excitations when we consider small
disturbances of the simple rotational states.

Let us clarify this point on the example of the meson-like string with massive
ends. For this model exotic states are rotations of $n$ times folded
rectilinear string. It was proved in Ref.~\cite{PeSh} that in
$3+1$\,-\,dimensional Minkowski space only these states
$$X^\mu(\tau,\sigma)=x_0^\mu+p^\mu\tau+\alpha_n\cos(\tilde\om_n\s+\phi_n)
(e_1^\mu\cos\tilde\om_n\tau+e_2^\mu\sin\tilde\om_n\tau),\quad
\sigma\in[0,\pi]$$
are motions of this string system with linearizable boundary conditions
(\ref{qq}), which take the linear form
$\big[\ddot X^\mu+(-1)^iQ_iX^{\prime\mu}\big] \Big|_{\sigma=\sigma_i}=0$
under restrictions
$\dot X^2\Big|_{\sigma=\sigma_i}=C_i^2={}$const.
Here $Q_i=C_i\gamma/m_i$ and $\tilde\om_n$ are roots of the equation
\be
(\tilde\om^2-Q_1Q_2)\big/\big[(Q_1+Q_2)\,\tilde\omega\big]=
\cot\pi\tilde\omega.
\label{zfreq}\ee

It is interesting that the same equation (\ref{zfreq}) describes physical
states --- excitations of the string with massive ends when we
consider small disturbances of its rotational motions (\ref{sol}). These
disturbances were studied in Refs.~\cite{AllenOV} but the results were
not correct. However in Ref.~\cite{stabPRD} these disturbances in the
linear approximation with using the conditions (\ref{ort}) were
obtained in the form
\begin{eqnarray}
X^\mu(\tau,\s)&=&X^\mu_{rot}(\tau,\s)+\!\sum_{n=-\infty}^\infty
\Big\{e_3^\mu\alpha_n\cos(\tilde\om_n\s+\phi_n)\exp(-i\tilde\om_n\tau)
\nonumber\\
&&{}+\beta_n\big[e_\parallel^\mu f_\parallel(\s)+i\big(e_0^\mu f_0(\s)+
e_\perp^\mu f_\perp(\s)\big)\big]\exp(-i\Omega_n\tau)\Big\}.
\label{osc}
\end{eqnarray}
Here $X^\mu_{rot}$ is the pure rotational motion (\ref{sol}),
$e_\parallel^\mu(\tau)=e_1^\mu\cos\tilde\om_1\tau+e_2^\mu\sin\tilde\om_1\tau$,
$e_\perp^\mu={\tilde\om_1}^{-1}\frac d{d\tau}e^\mu(\tau)$; $e_0$, $e_1$, $e_2$,
$e_3$ is the orthonormal tetrad. Each term in Eq.~(\ref{osc}) describes the
string oscillation that looks like the stationary wave with $n$ nodes.
There are two types of these stationary waves: (a) orthogonal oscillations
along $z$ or $e_3$-axis at the frequencies proportional to the roots
$\tilde\om_n$ of Eq.~(\ref{zfreq}), and (b) planar oscillations
(in the rotational plane). In the latter case the eigenfrequencies $\Omega_n$
result from the equation
$\displaystyle\frac{(\Omega^2-\tilde Q_1^2)(\Omega^2-\tilde Q_2^2)
-4Q_1Q_2\Omega^2}
{2\Omega\big[Q_1(\Omega^2-\tilde Q_2^2)+
Q_2(\Omega^2-\tilde Q_1^2)\big]}=\cot\pi\Omega,$
$\tilde Q_i^2=Q_i^2(1+v_i^{-2})$.
The frequencies $\Omega_n$ and $\tilde\omega_n$ are real numbers
so the rotations (\ref{sol}) of the string with massive ends are stable
in the linear approximation.

Let us make some conclusive remarks:
\begin{itemize}
\item Quasirotational motions of the string with massive ends in the
form of Fourier series (\ref{osc}) may be used as the basis of quantization
in the linear vicinity of the stable solution (\ref{sol}). Expression
(\ref{osc}) satisfies the constraint (\ref{ort}) so we have no an analog
of the Virasoro conditions in this case.
\item Progress in quantization of the considered nonlinear meson and baryon
string models is necessary for describing not only orbital but also radial
and other excitations of hadrons known from potential models
\cite{InSh}.
\item Quantization in the linear vicinity is possible only for stable
solutions, but we show that the rotational motions (\ref{sol}) for the linear
$q$-$q$-$q$ and three-string models are unstable on the classic level. However
this doesn't mean that these models are finally closed for further
applications.
\end{itemize}

\noindent{\bf Acknowledgements.}
{\small The work is supported by the RFBR grant 00-02-17359.}



\begin{thebibliography}{2}

\bibitem{BN}
B. M. Barbashov, V. V. Nesterenko, {\it Introduction to the
Relativistic String Theory}, World scientific, Singapore, 1990;
A. Chodos, C. B. Thorn, {\it Nucl. Phys. \bf B72} (1974) 509.

\bibitem{AY}
X. Artru, {\it Nucl. Phys. \bf B85}, 442 (1975);

\bibitem{Ko}
Yu.~I. Kobzarev, L.~A. Kondratyuk, B.~V. Martemyanov, M.~G. Shchepkin,
{\it Sov. J. Nucl. Phys. \bf 45} (1987) 330.

\bibitem{lin}
V. P. Petrov, G. S. Sharov,
{\it Mathem. Modelir., \bf 11} (1999) 39,
hep-ph/9812527.

\bibitem{CollinsPY}
P. A. Collins, J.L. Hopkinson, R.W. Tucker,
Nucl. Phys. {\bf B100} (1975) 157;
M.~S.~Plyushchay, G.~P.~Pronko, A.~V.~Razumov, {\it Theor. Math.
Phys. \bf63} (1985) 389;
{\it ibid.\ \bf 67} (1986) 576.

\bibitem{Tr}
G. S. Sharov, {\it Theor. Math. Phys. \bf 113} (1997) 1263;
{\it Phys. Rev. \bf D58} (1998) 114009;
hep-th/9808099.

\bibitem{stabPRD}
G. S. Sharov, {\it Phys. Rev. \bf D62} (2000) 094015, hep-ph/0004003.

\bibitem{4B}
G.~S.~Sharov, {\it Phys. Atom. Nucl. \bf 62} (1999) 1705,
hep-ph/9809465.

\bibitem{KalashCorn}
Yu. S. Kalashnikova, A.V. Nefediev, {\it Phys. Atom. Nucl. \bf 60} (1997) 1333.
J. M. Cornwall. {\it Phys. Rev. D. \bf54} (1996) 6527.

\bibitem{InSh} A. Inopin, G. S. Sharov, hep-ph/9905499.

\bibitem{PeSh}
V. P. Petrov, G. S. Sharov, {\it Theor. Math. Phys. \bf 109} (1996) 1388.

\bibitem{AllenOV}
T.~J.~Allen, M.~G.~Olsson, S.~Veseli,
{\it Phys.\ Rev.\ \bf D59} (1999) 094011;
{\it ibid. \bf D60} (1999) 074026, hep-ph/9903222.

\end{thebibliography}
\end{document}